\documentclass[12pt]{article}
\usepackage{amssymb,amsmath}
\usepackage{bm} %math bold symbols
\usepackage{booktabs} %some alignment stuff
\usepackage{array}
\usepackage{latexsym}
\usepackage{graphicx}
\usepackage{color}
\usepackage{datetime}
\usepackage[nosort]{cite}
\usepackage{verbatim}
\usepackage{enumerate}
\usepackage{chngpage} % allows for temporary adjustment of side margins
\usepackage{mathrsfs}
\usepackage{euscript}
\usepackage{psfrag}
\usepackage{subcaption}

\usepackage{datetime}

\usepackage[nosort]{cite}
\usepackage{chngpage} % allows for temporary adjustment of side margins
\usepackage{setspace}
\usepackage{tensor}

\usepackage[dvipsnames]{xcolor}

\newcommand{\be}{\begin{equation}}
	\newcommand{\ee}{\end{equation}}

\usepackage{tensor}

\usepackage{mciteplus}

\usepackage[colorlinks=true,      linkcolor=blue,      urlcolor=blue,      
filecolor=blue,      citecolor=blue,       pdfstartview=FitH,     
pdfpagemode=UseNone,      bookmarksopen=true]{hyperref}  % LINKS
\usepackage[all]{hypcap}     %should be loaded AFTER hyperref, which should otherwise be loaded last.

\definecolor{cardinal}{rgb}{0.6,0,0}
\definecolor{darkgreen}{rgb}{0,0.4,0}
\definecolor{golden}{rgb}{0.92, 0.7, 0}
\definecolor{midnight}{rgb}{0, 0, 0.5}
\definecolor{darkblue}{rgb}{0, 0, 0.7}
\definecolor{purple}{rgb}{0.5, 0, 0.5}

\def\IR{\mathbb{R}}

\numberwithin{equation}{section}

\begin{document}
	\begin{titlepage}

		\bigskip\bigskip\bigskip
		\bigskip
		\centerline{\LARGE \bf Born-Infeld Supermaze Waves}
		% D4-D2-P bound state from Born-Infeld action }
	\medskip
	\centerline{\LARGE \bf }

	\medskip

	\begin{center}

		\vspace{14mm}
		
		{\large
			\textsc{Iosif Bena and Rapha\"el Dulac}}
		\vspace{12mm}
		
		\textit{ Institut de Physique Th\'eorique, \\  
			Universit\'e Paris-Saclay, CNRS, CEA,\\
			Orme des Merisiers, Gif-sur-Yvette, 91191 CEDEX, France  \\}  
		\medskip
		
		\vspace{4mm} 
		%
		
		% Emails
		{\footnotesize\upshape\ttfamily  iosif.bena @ ipht.fr, raphael.dulac  @ ens.fr} 
		\vspace{13mm}

	\end{center}

	\bigskip
	\bigskip
	\bigskip
	\bigskip
	\bigskip
	\bigskip

	\centerline{Abstract}
	\bigskip
	
The entropy of the supersymmetric  D2-D4-P black hole comes at weak coupling from D2-brane strips stretched between parallel D4 branes and carrying momentum waves. We use the DBI action of D4 branes to construct two pieces of plumbing that enter in the construction of these microstates. The first is a semi-infinite D2 brane ending on a D4 brane and carrying a momentum wave along the common D2-D4 direction. The second is a non-Abelian solution to the 5D maximally-supersymmetric $SU(2)$ Super-Yang-Mills theory describing a momentum-carrying D2 strip stretched between two D4 branes. The solution without momentum is the same as the 't Hooft-Polyakov monopole, and the fields that carry the momentum can be added without changing any of the fields of the monopole.

\end{titlepage}
\tableofcontents
\section{Introduction}
A major success of String Theory is reproducing the Bekenstein-Hawking entropy of black holes. This is done by counting brane configurations that have the same charges as these black holes in a regime of parameters where gravity is turned off and relying on supersymmetry to argue that the number of states remains the same as one moves to the regime of parameters where gravity is finite and the classical textbook black hole exists \cite{strominger1996microscopic,maldacena1996black,dijkgraaf1997bps}. However, little is know of the fate of individual microstates as one moves to this regime. Some of these microstates have been shown to correspond to smooth horizonless microstate geometries, which have the same asymptotics as the black hole, but in which the black-hole horizon is replaced by as smooth cap \cite{bena2015habemus,Bena:2006kb,Bena:2016ypk}. 

However, even if these microstate geometries form the largest set of solutions to Einstein's equations ever constructed \cite{Heidmann:2019zws} their entropy is parametrically smaller than the Bekenstein-Hawking entropy of the black hole\footnote{For the D1-D5-P black hole, of entropy $2 \pi \sqrt{N_1 N_5 N_p}$, the entropy of smooth horizonless superstratum solutions is of order $\sqrt{N_1 N_5 \sqrt{N_p}}$.} \cite{mayerson2021counting}. To bypass this problem, one of the authors with Hampton, Houppe, Li and Toulikas proposed a method \cite{bena2023amazing} to track the microstates of the Type IIA F1-NS5-P system from the zero-gravity regime of parameters where these microstates are counted \cite{dijkgraaf1997bps} to the regime of parameters where the classical black hole exists, and to argue that the backreaction of the individual microstates of this system will give rise to horizonless solutions.

The M-theory uplift of a Type IIA F1-NS5 system consists of  $N_5$ M5 branes located at various positions on the M-theory circle, and $N_2$ M2 branes wrapping this circle. Each M2 brane can split into $N_5$ strips (corresponding to the Type IIA little strings), which can move independently. This gives rise to a complicated maze-like structure.  The resulting $N_2 N_5$ M2 strips can carry momentum independently, and their entropy reproduces exactly the entropy of the M2-M5-P black hole. However, as argued in \cite{bena2023amazing}, when taking into account interactions between branes, the story becomes a bit more complicated. The M2 branes pull on the M5 branes (similar to the way in which F1 strings terminating on a D3 brane pull on this brane \cite{callan1998brane}), resulting in a non-trivial spike-like structure times the line corresponding to the common M2-M5 direction. This configuration has 8 supercharges globally but as one zooms in near the spike one finds that the number of preserved supercharges is enhanced to 16 \cite{bena2023amazing}. Furthermore, if one adds momentum to this configuration, one can also argue that, even if only 4 supercharges are preserved globally, the supersymmetry is also enhanced to 16 supercharges locally near the brane profile.  As argued in \cite{bena2023amazing,Bena:2022fzf}, this local enhancement of supersymmetry is the hallmark of the existence of smooth horizonless solutions in the regime of parameters when the branes backreact \footnote{The intuition behind this argument is very simple: a black-hole horizon remains a black-hole horizon in any duality frame \cite{Horowitz:1996ay}, while a brane with 16 supercharges can be dualized to a D6 brane which uplifts in M-theory to the center of an empty Taub-NUT space.}.

The purpose of our paper is to construct explicitly the building blocks of the momentum-carrying supermaze. We do this by smearing the M2 branes of the supermaze along one of the directions of the four-torus wrapped by the M5 branes, and reduce the solution to Type IIA String Theory along this direction. The supermaze consists now of D2 strips extended between parallel D4 branes and carrying momentum along the common D2-D4 direction. One building block of the configuration without momentum is a semi-infinite D2 brane ending on and pulling the D4 brane. From a D4 brane perspective this configuration is described by the same solution to the DBI action as a semi-infinite D1-brane ending on a D3-brane \cite{Constable:1999ac}. In Section \ref{Bion with momentum} we present this solution and construct several new solutions in which this brane configuration carries momentum. We identify the various momentum waves as coming from 2-2 strings or 4-4 strings and we display the dipole charges that allow the supersymmetry to be enhanced locally to 16 supercharges.

We then consider a more complicated configuration, consisting of a D2 strip stretched between two parallel D4 branes. In the absence of momentum, this system of branes is described by a solution to the non-Abelian Born-Infeld action of two D4 branes. Fortunately, this solution is nothing but the 't Hooft-Polyakov monopole \cite{prasad35exact}, which describes a D1 string stretched between two D3 branes \cite{hashimoto1998shape,hashimoto2003strings}. Even if this monopole in constructed as a solution of the low-energy limit of this non-Abelian Born-Infeld action, which is the $SU(2)$, ${\cal N}=4$ Super-Yang-Mills theory theory, we know that it will also be a solution to the  parent non-Abelian Born-Infeld theory \cite{myers2000dielectric}.

We then add momentum to this configuration along the common D2-D4 direction. From the perspective of the D4-brane action, the oscillating momentum-carrying fields are electric and magnetic fields, which give rise to a nontrivial Poynting vector and do not change the profiles of the fields that give rise to the original momentum-less D4-D2 configuration. However, if one uplifts these branes to M-theory, one can see that some of the oscillating fields encode the angles between the M5 branes and the M-theory direction and between the M2 branes and the M-theory direction. Hence the unbent D4-D2 configuration with electric and magnetic fields that oscillate as a function of the common D2-D4 direction is uplifted to a M5-M2 configuration with a transverse wave carrying momentum, where the  transverse wave is polarized along the M-theory direction. This is precisely the configuration considered in \cite{bena2023amazing}.

Even if the local enhancement of the supersymmetry to 16 supercharges is hard to see from the perspective of the non-Abelian Born-Infeld action \cite{toappear2}, we show that the momentum-carrying wave  is localized at the intersection of the D2 and D4 branes, and that adding it does not affect locally any of the supersymmetries visible in the DBI description of the momentum-less D4-D2 configuration.

\section{A BIon furrow with momentum}\label{Bion with momentum}

To construct a Born-Infeld furrow, corresponding to a semi-infinite D2 brane ending on a string inside a D4 brane, it is easiest to review first the BIon  - an Abelian monopole solution which describes a semi-infinite D1 brane ending on a D3 brane \cite{Constable:1999ac}.

\subsection{The BIon monopole}
The BIon is a solution of Born-Infeld action:
\begin{equation}
	S_{BI}=-T_{3}\int \mathrm{d}^{4}\sigma \sqrt{-\mathrm{det}(g_{\mathrm{ij}} +F_{\mathrm{ij}})}\,,
\end{equation}
where $g_{ij}=g_{\mu \nu}\partial_iX^{\mu}\partial_jX^{\nu}$ is the pull back of the metric and $F_{ij}$ is the field strength, $T_{3}$ is the tension of the D3-brane. In spherical coordinates the BIon monopole is given by 
\begin{align}
	&\Phi=\frac{b}{r} \nonumber \\
	&F_{9r} \equiv \partial_r\Phi \label{BIon} \\
	&F_{\hat{\theta} \hat{\phi}}=F_{9r}\,. \nonumber
\end{align}
The second equation is the statement that one can either see this solution in a three-dimensional theory, describing a D1-brane spike on a D3, parameterized by the profile of the scalar \cite{Constable:1999ac} or as an equation describing the self-duality of the field strength in a four-dimensional theory\footnote{In the T-dual D4 worldvolume theory, this equation describes a distributions of D0 branes smeared along $x_9$, T-dual to the BIon.}. We used the conventions of \cite{callan1998brane} for normalizing $\theta, \phi$ as flat  coordinates: $\hat{\theta}, \hat{\phi}$. 
As an Abelian solution of the Born-Infeld action, this solution will preserve 16 supersymmetries at every location along the D3 worldvolume (more on this in Section \ref{supercharges-BIon}). The self-duality condition encoded in the last equation indicates that 8 of these supersymmetries are common to all locations, and hence will be preserved by the full solution. 

Indeed, the space-filling usual D3 brane condition already breaks half of the 32 vacuum SUSY, plus the extra condition:
\begin{align}
	&\hspace{12mm}F_{\mu \nu}\Gamma^{\mu\nu}\varepsilon=0 \label{susy condition} \\
	&\implies F_{r9}(\Gamma^{r9}+\Gamma^{\theta \phi})\varepsilon=0\,.
\end{align}
In the second equation the  $\Gamma$ matrices are flat (they square to the identity). 

The energy density of this solution per unit of D3-brane worldvolume diverges.  However, it is easy to understand that this divergence comes from the presence of the semi-infinite string ending on the D3 branes. Indeed we have: $\mathcal{H}=1+(\partial_r\Phi)^2$, and therefore as one integrates it over the D3-brane we obtain:
\begin{equation}\label{energy D1 brane}
	E=T_{3}\int \mathrm{d}^{3}r \frac{b^2}{r^{4}}=4\pi b T_{3}\int \mathrm{d}r \frac{b}{r^2}=4\pi b T_{3}\int d\Phi\,.
\end{equation}
In deriving the last equality we used the fact that $\mathrm{d}\Phi=\frac{b}{r^2}\mathrm{d}r$; this illustrates the fact that the energy density per unit length of the spike is constant and proportional to the tension of the D1 brane  \cite{Constable:1999ac, callan1998brane}. There is one further quantization for $b=N\pi$, with $N$ an integer, this gives exactly the D1-brane tension in \eqref{energy D1 brane}.

\subsection{The BIon furrow with momentum.}
\label{sec:BIon}

The BIon solution \eqref{BIon} is also a solution of the DBI action of a D4 action, where the D4 brane is extended along $y$, an isometry direction, as well as in an $\IR^3$ in which $r$ is a radial coordinate. This solution describes a semi-infinite D2 extending in the $y$ direction and pulling the D4 brane into a BIon furrow. Our purpose is to add a momentum wave along the $y$ direction while keeping the 16 local supersymmetries.

There are several ways in which one may try to add momentum to this furrow. The most obvious one is to add to $\Phi$ a momentum wave: $\Phi=\frac{b}{r}+f(y-t)$ (with $F_{\theta \phi}$ still given by the BIon value). This solution satisfies the DBI equations of motion, and is supersymmetric. It corresponds to a momentum wave carried by (4,4) open strings, that begin and end on the D4 brane. This momentum wave can be added to the D4 branes independently of the presence of the D2 brane, and its strength  does not depend on the position of the D2 branes. 

It is also possible to add momentum while maintaining the spherical symmetry of the BIon furrow, by giving a nontrivial profile to the D4 worldvolume electric fields $F_{ry}, F_{0y}$. This solution corresponds to (2,2) strings, and its structure is similar, but much simpler than to the non-Abelian wave we will construct in Section \ref{'t Hooft Pol section}. To still preserve the same supersymmetries as D4 branes, D2 branes and momentum along $y$, these momentum-carrying worldvolume fields must satisfy $F_{ry}=\pm F_{0r}$. One can see from Equation \eqref{susy condition} that the $\pm $ choice is related to the orientation of the momentum wave and its corresponding projector described in Appedix  \ref{dictionary}:
\begin{equation}
	F_{ry}\Gamma^{ry}\left(1\mp \Gamma^{0y}\right)\varepsilon=0\,.
\end{equation}

In the Abelian Born-Infeld theory, all the BPS solutions satisfy also the  linear electromagnetism equations of motions\cite{tseytlin2000born}. We first look for a solution to these equations:
\begin{align}
	&\mathrm{d}\star F=0\\
	&\mathrm{d}F=0\,.
\end{align}
And we find the solution:
\begin{align}
	&\Phi=\frac{b}{r}\\
	&\partial_r \Phi=F_{9r}=F_{\theta \phi}\\
	&F_{ry}=-F_{r0}=\frac{f(y-t)}{r^2}\,,
\end{align}
where $f(y-t)$ is an arbitrary function. The last condition can be realized if the gauge potential is $A_{0}=-A_y=\frac{f(y-t)}{r}$. 
The condition $F_{ry}=-F_{r0}$ which was imposed above to preserve supersymmetry, appears here as a consequence of the equations of motion and to avoid generating a nontrivial $F_{0y}$.

If $f(y-t)$ were a constant $c$, this solution would be T-dual to a fundamental string spike along the $y$ direction. Indeed we could identify $A_y$ with a transverse scalar, and we would have $\Phi=\frac{c}{r}$ coupled to a worldvolume electric field. This is the same solution as BIon, except that we turned on an electric field (like in the Callan-Maldacena spike) instead of the magnetic one. Hence, the T-dual solution of the solution with constant $f(y-t)$ describes a D3 brane with a D1 spike in the $x_9$ direction and an orthogonal F1 spike in the $y$ direction.

In the rest of the paper we will be interested in solutions which do not have this global charge, so we impose the extra condition:
\begin{equation}
	\int \mathrm{d}y f(y-t)=0\,.
\end{equation}
As expected, one can check that this solution satisfies the equations of motion of the full Born-Infeld theory.

\subsection{Energy analysis}
Let us now compute the energy of our configuration. The conserved charge associated to $F_{r0}$ is:
\begin{equation}
	\Pi_r=\frac{\partial \mathcal{L}}{\partial F_{r0}}=T_4\frac{F_{r0}\sqrt{r^4\sin(\theta)^2(1+F_{\theta\phi}^2)}}{\sqrt{1+F_{r9}^2-F_{r0}^2+F_{ry}^2}}\,.
\end{equation}

Therefore the Hamiltonian density $\mathcal{H}$ is given by:
\begin{align}
	\mathcal{H}&=T_4\frac{\sqrt{r^4\sin(\theta)^2(1+F_{\theta\phi}^2)}}{\sqrt{1+(\partial_r\Phi)^2-F_{r0}^2+F_{ry}^2}}\left(1+(\partial_r\Phi)^2+F_{ry}^2\right)\\
	&=T_{4}r^2\sin(\theta)(1+(\partial_r\Phi)^2+F_{ry}^2)\,.
\end{align}
By integrating the Hamiltonian density, $\mathcal{H}$, over the D4 worldvolume, we obtain its total energy $H$:
\begin{equation}\label{energetics}
	H=T_{4}\int d^{4}x (1+(\partial_r\Phi)^2+F_{ry}^2)=T_{4}\int d^{4}x+4\pi b T_{4}\int \mathrm{d}y\mathrm{d}\Phi \left(1+\frac{f(y-t)^2}{b}\right)\,,
\end{equation}
This equation reveals the BPS nature of our configuration and the fact that the energy is the sum of its three charges. Much like in the BIon, the constant in the bracket in \eqref{energetics} corresponds to the energy of the semi-infinite D2 brane ending on the D4 brane (recall the quantization $b=N\pi$). The second term is the local energy density of a momentum wave propagating in the $y$ direction.

\subsection{Who carries the momentum ?}

In our D2-D4 system the momentum can be carried in three ways: by (4,4) open strings (with both ends on the D4 brane), by (2,2) strings (with both ends on the D2 brane) or by (2,4) strings (with one end on the D2 brane and one end on the D4 brane)\cite{maldacena1996black}. 
In this subsection we show that the momentum of the solution we constructed above is carried by (2,2) strings. One intuitive way to see this is from the \emph{r.h.s} of Equation \eqref{energetics}, which contains an integral of the energy of the momentum wave over the entire D2-brane surface. This density does not diminish as one goes on the D2 spike away from the D4 branes, and hence it is unlikely that it will correspond to (2,4) strings, which are localized at the junction of the branes. 

We can also see this by expanding in different limits the $\kappa$-symmetry projection equation which gives the supersymmetries preserved by our branes:
\begin{align}
	\Gamma_{\kappa}&=\left(\frac{\Gamma^{0r\theta\phi y}i\sigma_2+\partial_r\Phi\Gamma^{09\theta\phi y}i\sigma_2}{1+(\partial_r\Phi)^2}\right)(1+\sigma_3F_{\theta\phi}\Gamma^{\theta\phi})\left(1+\sigma_3F_{yr}(-\Gamma^{0}+\Gamma^{y})\left(\frac{\Gamma^{r}+\partial_r\Phi\Gamma^9}{1+(\partial_r\Phi)^2}\right)\right)\,.
\end{align}
In the $r\rightarrow \infty$ limit this becomes the projection matrix of a D4 brane extended along the directions $r\theta\phi y$ and with no other charges:
\begin{equation}
	\Gamma_{\kappa \hspace{1mm} r\rightarrow \infty}=\Gamma^{0r\theta\phi y}i\sigma_2\,.
\end{equation}
In the opposite limit, $r\rightarrow 0$, we obtain:
\begin{equation}
	\Gamma_{\kappa \hspace{1mm}r\rightarrow 0}=\Gamma^{0y9}i\sigma_2\left(1+\sigma_3\frac{f(y-t)}{b}(\Gamma^{09}+\Gamma^{y9})\right)\,.
\end{equation}
This $\kappa$-symmetry projection is exactly the one of a D2-brane in the $y$-9 plane, with momentum along the $y$ direction. We conclude that the momentum is carried by (2,2) open strings.

As mentioned above, one can also add to this solution a momentum wave carried by (4,4) open strings, via a scalar-field wave: $\tilde{\Phi}=f(y-t)$. This momentum wave also also satisfies the equation of motion and preserves the 4 global supersymmetries, and can be added to the the solution independently of the (2,2) momentum wave. 
As explained in \cite{bena2023amazing}, when considering multiple D4 branes and D2 brane strips stretched between them, it is the (2,2) strings that carry the momentum of the strips, and that give the large number of states needed to reproduce the black-hole entropy. This being said, it would be quite interesting to try also to construct a wave on the BIon furrow that corresponds to momentum carried by (2,4) strings.

\section{The charges, dipole charges and supersymmetries of the BIon furrow with momentum.}\label{supercharges-BIon}

The purpose of this section is to read out the local charges and supersymmetries of the momentum-carrying BIon furrow, and to show that this solution is a themelion: it has 16 local supercharges and 4 global ones. We will also illustrate the relation between the dipole charges and those of the supermaze. We review the charges and dipole charges  of the BIon furrow and of a simple D2 brane solution carrying a momentum wave \cite{toappear} in Appendix \ref{appendix-charges}.

We first recap the field content of our solution:
\begin{align}
	&\Phi=\frac{b}{r}\\
	&\partial_r\Phi=F_{\theta \phi}\\
	&A_{y}=\frac{1}{r} f(y-t)\\
	&A_{t}=-\frac{1}{r} f(y-t)\\
	&F_{ry}=-F_{r0}=-\frac{f(y-t)}{r^2}\,.
\end{align}
The $\kappa$-symmetry projector that controls the Killing spinors of our solution is $1+\Gamma_{\kappa}$, where:
\begin{equation}
	\Gamma_{\kappa}=\left(\frac{\Gamma^{0r\theta\phi y}i\sigma_2+\partial_r\Phi\Gamma^{09\theta\phi y}i\sigma_2}{1+(\partial_r\Phi)^2}\right)(1+\sigma_3F_{\theta\phi}\Gamma^{\theta\phi})\left(1+\sigma_3F_{yr}(-\Gamma^{0}+\Gamma^{y})\left(\frac{\Gamma^{r}+\partial_r\Phi\Gamma^9}{1+(\partial_r\Phi)^2}\right)\right)\,.
\end{equation}
The first two brackets in this expression are exactly the $\kappa$-symmetry projector of the D4-D2 BIon without momentum, given in Equation \eqref{BIon kappa sym}. The full projection equation can be written as:
\begin{align}
	\left(1+\Gamma_{\kappa}\right)\varepsilon=0\,.
\end{align}
Upon factorizing by $\Gamma_{\kappa \mathrm{BIon}}$, the $\kappa$-symmetry projector of the BIon, this gives 
$0=(1+\Gamma_{\kappa})\varepsilon=(1+\Gamma_{\kappa \mathrm{BIon}}(...))\varepsilon \implies (\Gamma_{\kappa \mathrm{BIon}}+(...))\varepsilon=0$  as by definition $\Gamma_{\kappa \mathrm{BIon}}^2=1$. This leads to the projection equation:
\begin{align}
	\left(\left(\frac{\Gamma^{0r\theta\phi y}i\sigma_2+\partial_r\Phi\Gamma^{09\theta\phi y}i\sigma_2}{1+(\partial_r\Phi)^2}\right)(1+\sigma_3F_{\theta\phi}\Gamma^{\theta\phi})+1+\sigma_3F_{yr}(-\Gamma^{0}+\Gamma^{y})\left(\frac{\Gamma^{r}+\partial_r\Phi\Gamma^9}{1+(\partial_r\Phi)^2}\right)\right)\varepsilon=0\,.
\end{align}
By reordering the terms, defining the furrow tilt angle as $\tan(\alpha) \equiv \partial_r\Phi$ and by writing $1=\cos(\alpha)^2+\sin(\alpha)^2$ we obtain:
\begin{align*}
	&\cos(\alpha)\left(\cos(\alpha)+\sin(\alpha)\Gamma^{9r}\right)\left(1+\Gamma^{0r\theta\phi y}i\sigma_2\right)\varepsilon\\ \nonumber
	&+\sin(\alpha) \left(\sin(\alpha)-\cos(\alpha)\Gamma^{9r})(1+\Gamma^{09y}\sigma_1\right)\varepsilon\\ \nonumber
	&-\frac{F_{yr}}{\sqrt{1+(\partial_r\Phi)^2}}\sigma_3 \left(\cos(\alpha)\Gamma^{ry}+\sin(\alpha)\Gamma^{9y}\right)(\Gamma^{0y}+1)\varepsilon=0\,.
\end{align*}
It is clear from this expression that the solution we construct preserves 4-global supersymmetries, corresponding to D4 branes, D2 branes and momentum along the common D2-D4 direction.
One may wonder why the factor $\frac{F_{yr}}{1+(\partial_r\Phi)^2}$ has been split in the equation above as $\frac{F_{yr}}{\sqrt{1+(\partial_r\Phi)^2}}\frac{1}{\sqrt{1+(\partial_r\Phi)^2}}$, and why in the second term we chose to replace the radial derivative of $\Phi $ with $\tan (\alpha)$ but in the first term we did not? It is because we now define the {\em wingling angle} of the furrow, $\beta$, via:
\begin{equation} \label{wingling angle}
	\tan \beta \equiv \frac{F_{yr}}{\sqrt{1+(\partial_r\Phi)^2}}\,.
\end{equation}

Note that, because of the $r$ dependence of $F_{ry}=\partial_r\Phi\frac{f(y-t)}{b}$, our wingling angle is well defined. At this point it is unclear why we call this a ``wingling'' angle. The M-theory uplift will make this more clear. 

We now want to use the $\kappa$-symmetry projector to read off the dipole charges of our configurations. The relation between these charges and the supersymmetry projector was discussed in detail in \cite{Bena:2011uw, bena2023amazing} and to put the projector in the form that allows us to read off the charges we multiply it with:
\begin{equation}
	\cos(\beta) \left(\cos(\beta)+\sin(\beta)\sigma_3(\cos(\alpha)\Gamma^{ry}+\sin(\alpha)\Gamma^{9y})\right)\,.
\end{equation}
The full projection equation becomes:
\begin{align*}
	&\cos(\alpha)\left(\cos(\beta)^2\cos(\alpha)+\cos(\beta)^2\sin(\alpha)\Gamma^{9r}+\cos(\beta)\sin(\beta)\Gamma^{ry}\sigma_3\right)\left(1+\Gamma^{0r\theta\phi y}i\sigma_2\right)\varepsilon\\ \nonumber
	&+\sin(\alpha) \left(\sin(\alpha)\cos(\beta)^2-\cos(\alpha)\cos(\beta)^2\Gamma^{9r}+\cos(\beta)\sin(\beta)\Gamma^{9y}\sigma_3)(1+\Gamma^{09y}\sigma_1\right)\varepsilon\\ \nonumber
	&+ \sin(\beta)\left(\sin(\beta)-\cos(\beta)\cos(\alpha)\Gamma^{ry}\sigma_3-\cos(\beta)\sin(\alpha)\Gamma^{9y}\sigma_3\right)(\Gamma^{0y}+1)\varepsilon=0\,.
\end{align*}
All the terms which were odd in $\cos(\alpha)$ or $\sin(\alpha)$ cancel. When one develops the full projection equation, one can see that all the terms in which there is no $\Gamma^{0}$ also cancel.
Using the expressions  in Appendix \ref{dictionary} we can identify the global charges:
\begin{align}
	&Q^{\rm D4}_{\mathrm{r}\theta\phi\mathrm{y}}= M \cos(\alpha)^2\cos(\beta)^2\\
	&D^{\rm D2}_{\mathrm{9y}}=M \sin(\alpha)^2\cos(\beta)^2\\
	&Q^{\rm P}_{\mathrm{y}}=M \sin(\beta)^2 \,,
\end{align}
where $M$ is the energy density.
As expected, these charges correspond to D4 branes along $r\theta\phi y$, D2 branes along $9y$, and momentum along $y$. The local dipole charges of this configuration are:

\begin{equation}
	\begin{array}{ll}
		Q^{\rm D4}_{\mathrm{9}\theta\phi\mathrm{y}}=M \cos(\beta)^2 \cos(\alpha)\sin(\alpha)& Q^{\rm D2}_{\mathrm{ry}}=-M \cos(\beta)^2\cos(\alpha)\sin(\alpha)\\
		Q^{\rm D2}_{\theta\phi}=M \cos(\beta) \sin(\beta)\cos(\alpha)&Q^{\rm F1}_{ \mathrm{r}}=- M \cos(\beta)\sin(\beta)\cos(\alpha)\\
		Q^{\rm D0} =M \cos(\beta) \sin(\beta)\sin(\alpha)&Q^{\rm F1}_{ \mathrm{9}}=- M \cos(\beta)\sin(\beta)\sin(\alpha)\,.
	\end{array}
\end{equation}

So we conclude that locally, we have dipole charges corresponding to D4 branes along $9\theta \phi y$, D2 branes along $\theta \phi$ and along $ry$, D0 branes, as well as fundamental strings along $r$ and 9. Note that these charges can also be read off directly from the Born-Infeld and Wess-Zumino actions of the brane. 
Once again the BPS nature of this configuration shows up by the beautiful factorization:
\begin{align}
	&\sum_i Q_{i,\mathrm{global}}=M\\
	&\sum_i Q_{i}^2=M^2\,.
\end{align}
We recall the definition of the angles before giving the expression of the global charges:
\begin{align}
	&\tan(\alpha)\equiv \partial_r\Phi\\
	&\tan(\beta) \equiv \frac{\partial_r\Phi}{\sqrt{1+(\partial_r\Phi)^2}}f(y-t)\,.
\end{align}
The ratio between the momentum and the total energy density is:
\begin{equation}
	\sin(\beta)^2=\frac{F_{ry}^2}{1+(\partial_r\Phi)^2+F_{ry}^2}=\frac{Q_P}{\mathcal{H}}\,.
\end{equation}
Here $F_{ry}^2=F_{0r} F_{ry}$ is the Poynting vector which carries the momentum, and the \emph{Hamiltonian} density $\mathcal{H}=1+(\partial_r\Phi)^2+F_{ry}^2$. We also obtain the proportion of the total energy to which the D4 and D2 branes contribute:
\begin{align}
	&\frac{Q_{D4}}{\mathcal{H}}=\cos(\alpha)^2\cos(\beta)^2=\frac{1}{1+F_{ry}^2+(\partial_r\Phi)^2}\\
	&\frac{Q_{D2}}{\mathcal{H}}=\sin(\alpha)^2\cos(\beta)^2=\frac{(\partial_r\Phi)^2}{1+F_{ry}^2+(\partial_r\Phi)^2}\,.
\end{align}

Figure 1 summarizes the global and local charges of our system in the diagrammatic language introduced in \cite{bena2023amazing}. This shows that our solution is a piece of a supermaze. 
\begin{figure}[h]
	\begin{subfigure}[h!]{0.35\linewidth}
		\includegraphics[width=\linewidth]{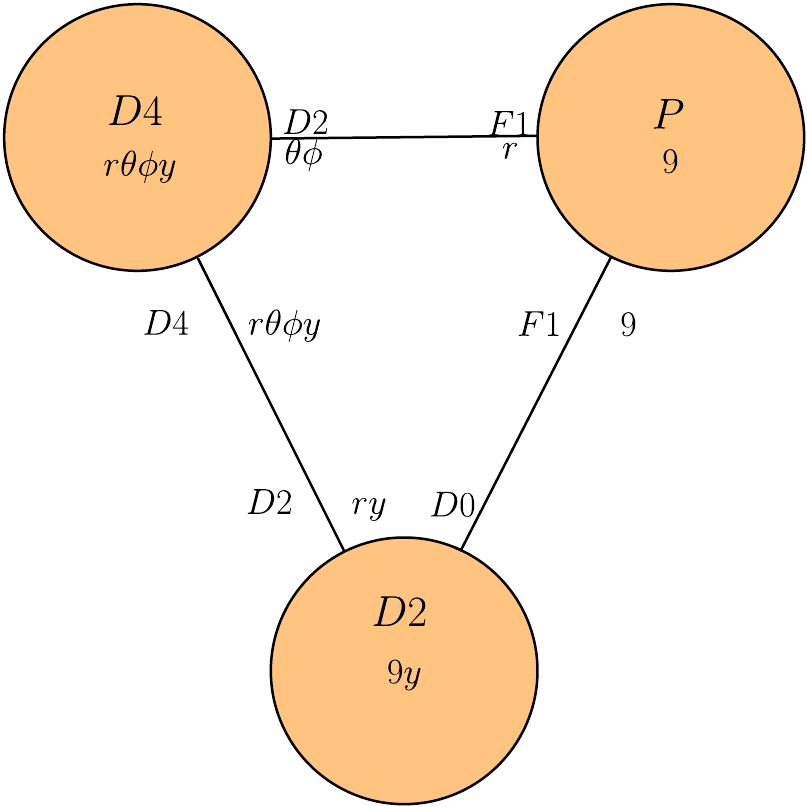}
		\caption{Triality in Type IIA}
	\end{subfigure}
	\hfill \label{triality2}
	\begin{subfigure}[h!]{0.35\linewidth}
		\includegraphics[width=\linewidth]{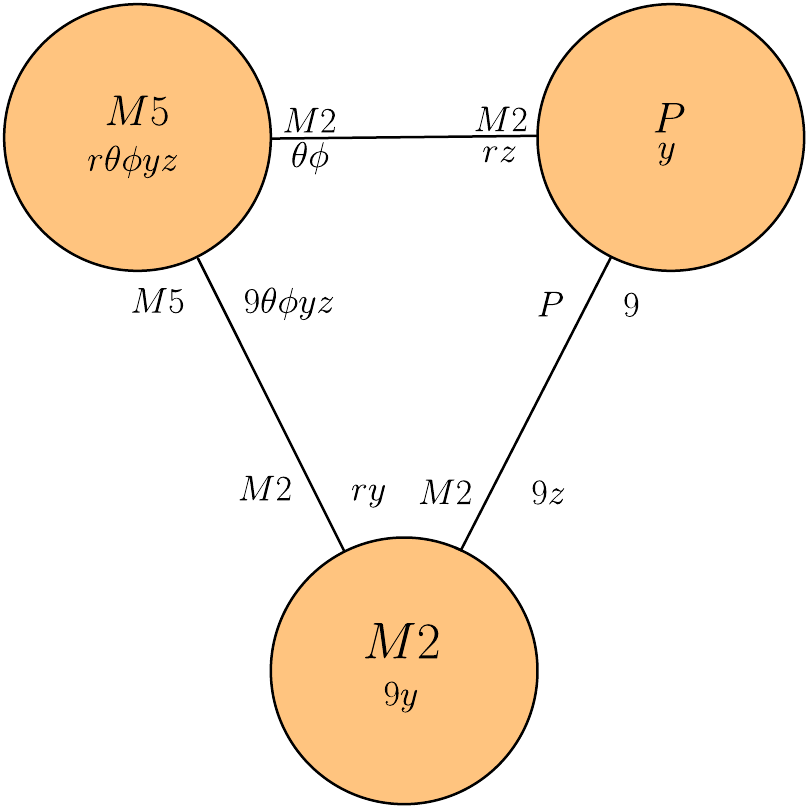}
		\caption{Triality in M theory}
	\end{subfigure}%
	\caption{Global and local charges of our solution in Type IIA String Theory (a) and M theory (b). The brane charges in the circles are global charges. The branes on the branches are the ``glue'' one needs to add in order to have enhance the local supersymmetry to 16 supercharge. The two ``glue'' branes on a given branch have the same local charge.}
\end{figure}

On the \emph{r.h.s} we present the uplift to M-theory of our solution, corresponding to an M5-M2-P bound state with local M2 and M5 brane dipole charges. The wingling angle is the angle at which the  M2 and M5 branes bend in order to carry momentum.

\section{Adding momentum to a D2 brane stretched between two D4 branes
	\label{'t Hooft Pol section}}

Having succeed to put a momentum wave on a semi-infinite D2 brane ending on a D4 brane, we now turn to the more complicated goal of putting a momentum wave on a D2-brane strip stretched between two D4 branes. As we explained in the Introduction, this strip is the key building block of the supermaze, which was shown in \cite{bena2023amazing} to be able to reproduce the entropy of the Type IIA F1-NS5-P black hole. To describe the D2 strip we need to use the $SU(2)$ Non-Abelian Born-Infeld action. The D2 strip without momentum is given by the 't Hooft Polyakov monopole in the D4-brane worldvolume $\IR^3$ orthogonal to the D2 branes. We begin by reviewing this solution, and then add momentum along the common D2-D4 direction.

\subsection{The 't Hooft Polyakov monopole}

Since the non-Abelian generalization of the Born-Infeld action \cite{tseytlin1997non} is ambiguous \cite{bergshoeff2001non,kabat1998linearized}, we work in the $SU(2)$ Super-Yang-Mills theory which is the expansion of this theory for small electric fields. However, since the supersymmetric solutions of the Abelian Super-Yang-Mills theory are also solutions of the non-linear BI theory \cite{tseytlin2000born}, it is also possible that the solution we consider will be a solution of the non-Abelian Born-Infeld action.

The bosonic action of the Super-Yang-Mills theory in 4 dimensions is:
\begin{equation}
	S=-T_{4}\int\mathrm{d}^4x \frac{1}{4}(F_{\mu \nu}^a)^2+\frac{1}{2}(D_{\mu}\Phi^{a})^2\,,
\end{equation}
with 
\begin{equation}
	F_{\mu\nu}^{a} =\partial_{\mu}A_{\nu}^{a}-\partial_{\nu}A_{\mu}^{a}+\epsilon^{abc}A_{\mu}^bA_{\nu}^c, \hspace{20mm} D_{\mu}\Phi^{a}=\partial_{\mu}\Phi^{a}+\epsilon^{abc}A_{\mu}^b\Phi^c\,.
\end{equation}
The 't Hooft Polyakov monopole is a solution of the 4D Yang-Mills theory:
\begin{align}
	&A_{i}^{a}=\epsilon_{aij}\frac{x^{j}}{r}\frac{(1-K(r))}{r}\label{monopole1} \\
	&\Phi^{a}=\frac{x^{a}}{r^2}H(r) \label{monopole2}\,,
\end{align}
with $K(r)=\frac{Pr}{\sinh(Pr)},\hspace{5mm} H(r)=Pr\coth(Pr)-1$. 

From the perspective of the $SU(2)$ theory describing two D3 branes, this solution can be interpreted as describing a D1 brane stretching between these D3 branes \cite{hashimoto2003strings}. However, one can also see the 't Hooft Polyakov monopole  as a $y$-independent solution of the $SU(2)$ five-dimensional Super-Yang-Mills theory describing two D4 branes. This solution describes a D2 strip stretched between these D4 branes. 

This solution describes a non-diagonalizable solution to the $SU(2)$ non-Abelian theory. Hence, is very hard to distinguish the positions of the individual D4 branes except to show that at infinity they are at a fixed distance away from which other and that, as expected, this distance decreases as one approaches the location of the monopole. One can try to diagonalize the scalar field and show that its eigenvalues are  $\pm H(r)$\footnote{A plot is given in Figure 8 of \cite{hashimoto2003strings}.}; however, interpreting these eigenvalues as the positions of the D4 branes is not very straightforward, since the other fields with non-Abelian profiles cannot be simultaneously diagonalized.

\subsection{Adding momentum}

Much like for the Abelian BIon, the easiest momentum wave one can add corresponds to momentum carried by the (4,4) strings:  $\Phi=f(y-t)\mathrm{I}_{\mathrm{d}}$, where $f(y-t)$ is an arbitrary function and $\mathrm{I}_{\mathrm{d}}$ is the identity matrix. This wave can be added irrespective of the presence of a D2 strip. However, in order to describe a piece of the supermaze, we need to add momentum carried by the D2 strip; this would corresponds to an arbitrary null-wave profile of the worldvolume gauge fields. Inspired by the Abelian solution, we consider an ansatz:  $A_{y}=-A_{0}\sim f(y-t)$, with $f$ an arbitrary function, and do not change the profile of the other fields in the monopole solution (\ref{monopole1},\ref{monopole2}).

The 't Hooft Polyakov monopole breaks the $SU(2)$ gauge symmetry to a $U(1)$ through Higgs mechanism. Therefore, looking at this solution in field theory language, the spectrum is composed of 1 photon (massless gauge field) and 2 massive W bosons. If we want to add a gauge-field momentum wave, we have to use the {\em massless} components, and therefore the gauge fields should be along the same group element as the scalar field. The ansatz we will start with is therefore that of the 't Hooft Polyakov monopole in $r,\theta,\phi$, with two extra fields:
\begin{align}
	&A_{0}^{a}=\frac{x^{a}}{r}g(r) f(y-t)\\
	&A_{y}^{a}=-\frac{x^{a}}{r}g(r) f(y-t)\,,
\end{align}
where $g$ is an unknown function that we want to find. As mentioned before, the two gauge fields are proportional to the scalar field.  It is also straightforward to see that $F_{0i}=-F_{yi}$.
As in the Abelian solution, in order to preserve supersymmetry,  $F_{0y}$ should vanish:
\begin{equation}
	F_{0y}^{a}=\partial_0A_{y}^{a}-\partial_{y}A_{0}^{a}+\epsilon^{abc}A_{y}^{c}A_{0}^{b}=0\,.
\end{equation}
The first two terms cancel each other by construction. The last term vanishes because the two gauge-fields are proportional to each other and therefore commute. We now compute $F_{0i}^{a}$ ($F_{yi}^{a}$ can be computed in the same way):
\begin{align}
	F_{i0}^{a}&=\partial_iA_0^{a}+\epsilon^{abc}A_i^{b}A_0^{c}\\
	&=\frac{x^{i}x^{a}}{r^2}\left(\left(\frac{g}{r}\right)'-\frac{1-K}{r}g(r)\right)f(y-t)+\delta_{ia}\left(\frac{g}{r}+\frac{1-K}{r}g(r)\right)f(y-t)\,.
\end{align}
In the following we will denote $F_{9i}\equiv \mathrm{D}_{i}\Phi$, such that the equations of motion can be expressed in terms of field strengths:
\begin{equation}
	\mathrm{D}_{\mu}F^{\mu\nu}=0\,.
\end{equation}
Indeed, for $\nu=i$, this reduces to the equation of motion for the 't Hooft Polyakov monopole.
\begin{align}
	\mathrm{D}_{\mu}F^{\mu i}&=\mathrm{D}_{j}F^{ji}+\mathrm{D}_{9}F^{9i}+\mathrm{D}_{0}F^{0i}+\mathrm{D}_{y}F^{yi}\\
	&=\mathrm{D}_{j}F^{ji} +\mathrm{D}_{9}F^{9i}-\partial_0F_{0i}+\partial_{y}F_{yi}-[A_{0},F_{0i}]+[A_{y},F_{yi}]\,.
\end{align}
Since $\partial_0F_{0i}=\partial_{y}F_{yi}$ and $A_0=-A_y, \hspace{2mm}F_{0i}=F_{yi}$, the four last terms cancel each other. The equations of motion reduce to :
\begin{equation}
	\mathrm{D}_{j}F^{ji} +\mathrm{D}_{9}F^{9i}=0\,,
\end{equation}
which are the \emph{e.o.m} satisfied by the gauge fields of the 't Hooft Polyakov monopole. The equations of motion for the scalar field are:
\begin{equation}\label{scalar field equation of motion}
	\mathrm{D}_{i}F^{i9}=0\,.
\end{equation}
These equations are solved by the fields in Equations (\ref{monopole1},\ref{monopole2}), which satisfy the condition $\frac{1}{2}\epsilon_{ijk}F^{jk}=\mathrm{D}_{i}\Phi$ . This can be viewed as a self-duality condition with $F_{9i}=D_{i}\Phi$. Therefore the equation of motion for $A_0,A_y$ are:
\begin{equation}
	\mathrm{D}_{i}F^{iy}=0, \hspace{10mm} \mathrm{D}_{i}F^{i0}=0\,.
\end{equation}
As there is no derivative along $y$ and $0$, we can factorize in this expression an arbitrary function, $f(y-t)$, representing the profile of the null wave. The equations of motion for $\tilde{A}^{a}_{y}=-\tilde{A}^{a}_{0}=\frac{x^{a}}{r}g(r)$ reduce to an equation similar to the scalar-field equation of motion \eqref{scalar field equation of motion}. Therefore, the solution for $A_0,A_y$ is:
\begin{align}
	&A_{0}^{a}=\Phi^{a}f(y-t)\\
	&A_{y}^{a}=-\Phi^{a}f(y-t)\,.
\end{align}
Hence, the function $g(r)$ we were looking for is: $g(r)=H(r)$. 

It is not hard to see that this solution describes a null wave on the D2 brane strip, as it can only be built on top of an existing monopole solution. This is in contrast to the solution describing (4,4) strings,  $\Phi=f(y-t)\mathrm{I}_{\mathrm{d}}$, which can be added to the D4-brane theory regardless of the presence of a monopole. Had we tried to build a solution of our type in the absence of a monopole, using the ansatz $A_{y}^{a}=-A_{0}^{a}=\frac{x^{a}}{r}g(r)f(y-t)$ we would obtain $g=1$ and therefore:
\begin{equation}
	A_y^{a}=-A_{0}^{a}=\frac{x^{a}}{r}f(y-t)\,.
\end{equation}
This solution is actually pure gauge as it can be diagonalized to $A_y^{a}=f(y-t)\sigma_3$ and the field strength would be trivial. This argument also explains why the momentum wave is localized to  region of the monopole: at infinity the momentum wave becomes pure gauge and hence it has vanishing energy.

Just as in the Abelian solution, the momentum-carrying wave does not distort the shape of the fields that give rise to the momentum-less 't Hooft Polyakov monopole. From the perspective of the D4-brane theory, the momentum is carried by the interaction of the magnetic and electric fields, which give rise to a nontrivial Poynting vector.

\subsection{The supersymmetries of the solution}
As this solution is purely bosonic, the variation of the fermionic fields under supersymmetry must be trivial, which leads to
\begin{equation}
	F_{\mu\nu}^{a} \Gamma^{\mu\nu}\varepsilon=0\,. \label{fermions}
\end{equation}
The 't Hooft Polyakov monopole describes a D2 brane that is mutually BPS with respect to the D4 branes, Hence, the Killing spinors are ``killed'' by the space-filling D4-brane projector:
\be
(1-\Gamma^{0123y} i\sigma_2) \varepsilon=0\,.
\ee
The effect of the nontrivial self-dual field strengths, $F_{ij}\Gamma^{ij}+\varepsilon_{ijk}F_{k9}\Gamma^{k9}$, is to generate another projection condition: 
\be
\Gamma^{9123}\varepsilon=\varepsilon\,.
\ee
The second condition combined with the first one imply that the Killing spinors are also killed by the projector corresponding to a D2-brane along $9y$. In addition to those two conditions we also have the condition:
\begin{equation}
	(F_{iy}\Gamma^{iy}+F_{i0}\Gamma^{i0})\varepsilon=0\implies (1-\Gamma^{0y})\varepsilon=0\,.
\end{equation}
This is the projection equation corresponding to a  momentum wave traveling along the $y$ direction. This confirms that this solution is a BPS solution and should therefore be a solution of the full non-Abelian Born-Infeld action.

\subsection{How is the momentum carried?}

One can calculate the energy density of the full solution. In the absence of momentum, for the 't Hooft Polyakov monopole, this energy has been shown in \cite{hashimoto1998shape} to be proportional to the distance between the D4 brane times the tension of the D2 string stretching between them: $E\sim T\Delta X$. We can also compute the contribution to the total energy of the momentum  wave we have added:

\be
	\begin{aligned}
	E_{\rm \Delta P}&=T\int \mathrm{d}r\mathrm{d}\Omega_2\mathrm{d}y\, r^2 \, \rho_{\rm \Delta P} = T\int \mathrm{d}r\mathrm{d}\Omega_2\mathrm{d}y \, r^2
	\frac{1}{2}\left(F_{0i}^{a}F_{0}^{ia}+F_{yi}^{a}F_{y}^{ia}\right) \\
	&=4\pi T\int \mathrm{d}r\mathrm{d}y\hspace{1mm} r^2 \left(\left(\partial_r\left(\frac{H}{r}\right)\right)^2+2\frac{K^2H^2}{r^4}\right) f(y-t)^2\,. 
\end{aligned}
\ee
First, one can take the $P\rightarrow \infty$ limit, and the energy reduces to:
\begin{equation}
	E=4\pi T \int \mathrm{d}r\mathrm{d}y\frac{1}{r^2} f(y-t)^2\,.
\end{equation}
This is exactly the energy we found for the BIon in Equation \eqref{energetics} ! 

\begin{figure}[h]
	\centering
	\includegraphics[scale=0.55]{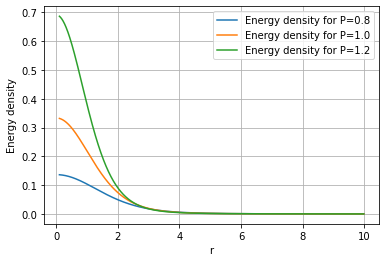}
	\caption{The energy density of the momentum wave, $\rho_{\rm \Delta P}$  as a function of $r$. One can see that the maximum of the energy density is localized at $r=0$, and is finite when $P$ is finite. The maximum value of the energy density scales like $P^4$. One can see that as the distance between the two D4 brane increases, the momentum becomes more localized around $r=0$.}
	\label{momentum energy}
\end{figure}

When the distance between the two D4 branes is finite, the energy density brought about by the momentum wave is localized near $r=0$ and is finite. In Figure \ref{momentum energy} we plot this energy density as a function of $r$ for different values of the distance between the D4 branes. One can see that as this distance becomes infinite, the energy density of the momentum wave diverges at the origin, reproducing the physics of the Abelian BIon furrow with momentum constructed in Section \ref{sec:BIon}.

\section{Conclusions and Future Directions}

We have constructed two solutions of the the Abelian and non-Abelian Born Infeld actions that describe pieces of the supermaze whose counting reproduces the entropy of the D2-D4-P black hole \cite{bena2023amazing}. The first is a semi-infinite D2 brane ending on a D4 brane and carrying momentum on the common D2-D4 direction. From a Born-Infeld-action perspective, the D2-D4 solution without momentum is the same as the BIon solution describing a D1 brane ending on a D3 brane \cite{Constable:1999ac}. We have added the momentum via worldvolume electric and magnetic fields, which, much to our surprise, do not change at all the shape of the other fields appearing in the BIon solution. We have also displayed the ``themelion'' structure of this solution \cite{bena2023amazing,Bena:2022fzf}: the fact that it has 16 supersymmetries everywhere locally, but only four of those (corresponding to to the common supersymmetries of the D2-D4-P system) are preserved everywhere. 

The second solution we constructed describes a D2-brane strip stretched between two D4 branes and carrying a momentum wave. In the absence of momentum, this solution is given  by a 't Hooft-Polyakov monopole \cite{prasad35exact} solution of the $SU(2)$ Super-Yang-Mills theory describing the two D4 branes. This solution involves only three of the D4 worldvolume directions, and is independent of the fourth, which is the common D2-D4 direction. 

We have been able to add a supersymmetric momentum wave to this non-Abelian 't Hooft-Polyakov monopole. The momentum is carried by nontrivial fluctuations of two of the worldvolume gauge fields, which do not commute with the other fields of the 't Hooft-Polyakov monopole. Despite this, adding the momentum wave does not affect any of these fields.  We have also shown that this non-Abelian null wave preserves the same supersymmetries as those of the D2-D4-P system.

One missing part of our investigation is ascertaining whether one can ascribe a {\em 16-local-supercharge} themelion structure to the non-commutative solution we found. This is a complicated question, because it is hard to define what one means by local in a non-Abelian solution. A few steps in this direction will appear in \cite{toappear2}, and we hope that the methods developed there will allow us to address this question for our solution as well.

Another open problem is to construct null momentum waves on more complicated monopole solutions, such as an $SU(3)$ monopole describing two D2 brane strips stretched between three D4 branes. One expects to be able to put independent momentum waves on each of these strips, and hence one expects the most general solution to be parameterized in terms of {\em two} arbitrary functions of $y-t$. It would be interesting to see whether these momentum waves also preserve the fields of the original momentum-less monopole. 

Last, but not least, we believe our work heralds some very good news for programme of constructing supergravity supermaze solutions and thus reproducing the full black-hole entropy via horizonless supergravity solutions \cite{Bena:2022rna}. This is because both for the Abelian and for the non-Abelian D4-D2 solutions, adding momentum using our specific combination of worldvolume electric and magnetic fields does not modify any of the other fields of the momentum-less solution. This hints at the presence of a linear structure underlying the supermaze supergravity solutions, similar to the ones found in the five- and six-dimensional supergravity ansatze underlying the known microstate geometries \cite{bena2005one,bena2012supersymmetric}. A hint of the existence of such a structure has already appeared in  \cite{bena2023maze}, and it would be interesting to see if such a linear structure exists, and whether it is also visible in the warped AdS$_3$ solutions describing near-horizon horizon geometries of the momentum-less supermaze \cite{Bachas:2013vza}.

\vspace{1em}\noindent {\bf Acknowledgements:} 
We would like to thank Nejc \v Ceplak, Bin Guo, Shaun Hampton, Yixuan Li, Dimitrios Toulikas and Nick Warner for interesting discussions. The work of IB was supported in part by the ERC Grants {\em 787320 - QBH Structure} and {\em 772408 - Stringlandscape}. 

\appendix 

\section{The Brane Projectors }\label{dictionary}

In this Appendix we review the projectors corresponding to various branes, strings and waves \cite{Bena:2011uw}. The F1 and momentum projectors are the same in Type IIA and Type IIB String Theory.

\begin{equation}\nonumber
	\begin{array}{l|l}
		P_{\rm P}=\Gamma^{01} & P_{\rm  F1}=\Gamma^{01}\sigma_3\\ 
		P_{\rm D0}=\Gamma^{0}i\sigma_2 & P_{\rm D1}=\Gamma^{01}\sigma_1\\
		P_{\rm D2} =\Gamma^{012}\sigma_1 & P_{\rm D3}=\Gamma^{0123}i\sigma_2\\
		P_{\rm D4}=\Gamma^{01234}i\sigma_2 &P_{\rm  D5}=\Gamma^{012345}\sigma_1\\
		P_{\rm D6}=\Gamma^{0123456}\sigma_1&P_{\rm D7}=\Gamma^{01234567}i\sigma_2
	\end{array}
\end{equation}

\section{The charges and dipole charges of  8-supercharge solutions}
\label{appendix-charges}

In this Appendix we explain how to read off the local dipole charges and the global charges from the 
$\kappa$-symmetry projector of certain branes with fluxes. As we explain in more detail in \cite{toappear}, the Killing spinors of a brane configurations can be constructed either in a {\em Hamiltonian} formalism, where one uses the conserved local charges of the solution, or in a {\em Lagrangian}, $\kappa$-symmetry formalism, which uses the electric fields and velocities of the solution, but not the conserved charges. The two descriptions are equivalent, and one can use the  $\kappa$-symmetry formalism to read of the local charges of the solutions we build and to show that they have the ``themelion'' structure: they preserve 16 supercharges locally, and a fraction of these are preserved by the global solution. 

In this Appendix we do this calculation for the eight-supercharge D2-D4 BIon furrow and for an eight-supercharge null momentum wave on a D2 brane. These D2-D4 and D2-P configurations are limits of the the four-supercharge D2-D4-P BIon furrow with momentum that we analyze in Section \ref{supercharges-BIon}.

\subsection{Charges and dipole charges of the BIon furrow}

In this subsection we explain how to read off the dipole charge of the BIon furrow directly from the  $\kappa$-symmetry projector. The Hamiltonian density is:
\begin{equation}
	\mathcal{H}=T_4(1+(\partial_r\Phi)^2).
\end{equation}
where $T_4$ is the D4 brane tension. Hence, the energy density is equal to the sum of the D4  charge density and the D2 charge density, $T_4 (\partial_r\Phi)^2$. The  $\kappa$-symmetry projection equation is 
\be
(1+\Gamma_{\kappa})\varepsilon=0 \,, \label{projector-general}
\ee
where the BIon $\Gamma_{\kappa}$ is:
\begin{align}\label{BIon kappa sym}
	\Gamma_{\kappa}&= \frac{\Gamma^{0r\theta\phi y}i\sigma_2+\partial_r\Phi\Gamma^{09\theta\phi y}i\sigma_2}{\sqrt{(1+F_{r9}^2)^2}}\left(1+F_{\theta\phi}\sigma_3\Gamma^{\theta\phi}\right).\\
	&=\frac{1}{1+(\partial_r\Phi)^2}\Gamma^{0r\theta\phi y}i\sigma_2+ \frac{\partial_r\Phi}{1+(\partial_r\Phi)^2}\Gamma^{09\theta\phi y}i\sigma_2-\frac{\partial_r\Phi}{1+(\partial_r\Phi)^2}\Gamma^{0ry}\sigma_1+\frac{(\partial_r\Phi)^2}{1+(\partial_r\Phi)^2}\Gamma^{0y9}\sigma_1.
\end{align}
Upon defining $\tan(\alpha) \equiv \partial_{r}\Phi$, this becomes:
\begin{equation}
	\Gamma_{\kappa}=\cos(\alpha)^2\Gamma^{0r\theta\phi y}i\sigma_2+\sin(\alpha)\cos(\alpha)(\Gamma^{09\theta\phi y}i\sigma_2-\Gamma^{0ry}\sigma_1)+\sin(\alpha)^2\Gamma^{0y9}\sigma_1.
\end{equation}

\begin{figure}[h]
	\centering
	\includegraphics[scale=0.6]{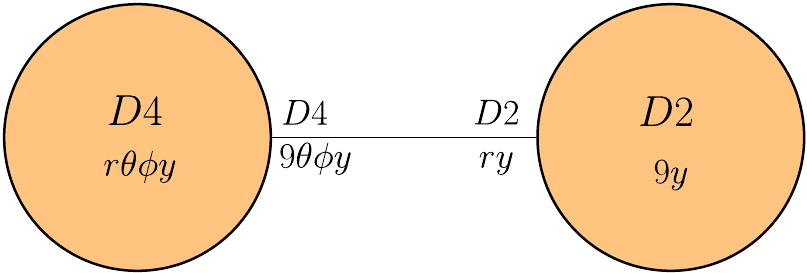}
	\caption{The branes in the circles correspond to the global charges. The branes on the branch are the ``glue'' one needs to  add in order to enhance the supersymmetry to 16 local supercharges.}
	\label{fig:2-charge-figure}
\end{figure}

As explained in \cite{bena2023amazing, Bena:2022fzf}, the equal and opposite dipole charges (corresponding to D2 branes along $ry$ and D4 branes along $9\theta\phi y$) act as a glue responsible for enhancing the local supersymmetry to 16 supercharges while preserving the 8 global supersymmetries of the original D2 branes and D4 branes. Using the dictionary in Appendix \ref{dictionary} we can read off the global and local (dipole) charges:

\begin{align}
	&Q^{\rm D4 \hspace{1mm}\mathrm{global}}_{\hspace{1mm} r\theta\phi y}=M\cos(\alpha)^2\\
	&Q^{\rm D4\hspace{1mm} \mathrm{local}}_{\hspace{1mm}9\theta\phi y}=M\cos(\alpha)\sin(\alpha)\\
	&Q^{\rm D2\hspace{1mm} \mathrm{local}}_{\hspace{1mm}ry}=-M\cos(\alpha)\sin(\alpha)\\
	&Q^{\rm D2 \hspace{1mm}\mathrm{global}}_{\hspace{1mm} 9 y}=M\sin(\alpha)^2,
\end{align}
where $M$ is the energy density of this solution. One can represent this system via the diagram in Figure \ref{fig:2-charge-figure}.

We can see that the energy of this configuration is equal to the sum of the global charges, as one expects in a solution preserving the 8-supercharges corresponding to these charges:
\begin{equation}
	Q_{\rm D2 \hspace{1mm}\mathrm{global}}+Q_{\rm D4 \hspace{1mm}\mathrm{global}} 
	=M(\cos(\alpha)^2+\sin(\alpha)^2)=M. \label{sum-charges}
\end{equation}
Furthermore, as in all systems that have 16 supercharges locally, the sum of the squares of all charges, local and global, is equal to the square of the energy density: 
\begin{equation}
Q^2_{\rm D2 \hspace{1mm}\mathrm{global}}+Q^2_{\rm D4 \hspace{1mm}\mathrm{global}} +
Q^2_{\rm D2 \hspace{1mm}\mathrm{local}}+Q^2_{\rm D4 \hspace{1mm}\mathrm{local}} 
=M^2(\cos(\alpha)^2+\sin(\alpha)^2)^2=M^2. \label{sum-charges-square}
\end{equation}
As we will see in Section \ref{supercharges-BIon}, these two properties will also be present in the BIon furrow with momentum.

\subsection{Charges and dipole charges of the D2-P system}

The second eight-supercharge system we want to discuss is the D2 brane with momentum along an arbitrary worldvolume direction, which we label $x_2$. The D2 brane can either carry the momentum via a null scalar wave, $\Phi=f(x_2-t)$, or via a worldvolume combination of the electric and magnetic fields $F_{01}=F_{12}=f(x_2-t)$. From a D2 brane perspective the two modes are rather different - one changes the shape of the brane and the other does not. However, upon uplifting to M-theory it is not hard to see that the two types of momentum modes are equivalent.

We will focus on the electric-magnetic worldvolume wave, where the momentum comes from a Poynting vector. The $\kappa$-symmetry projector \eqref{projector-general} has 
\begin{align}
	\Gamma_{\kappa}^{D2-P}&=\Gamma^{012}\sigma_1\left(1+F_{12}\sigma_3\left(\Gamma^{02}+\Gamma^{12}\right)\right).
\end{align}
where we used the fact that the on-shell Lagrangian, $\sqrt{1-F_{0r}^2+F_{yr}^2}$ is equal to one. Upon multiplying with $\Gamma^{012}\sigma_1$ the full projection equation is:
\begin{align}
	&\left( \Gamma^{012}\sigma_1+1+F_{12}\sigma_3\left(\Gamma^{02}+\Gamma^{12}\right)\right)\varepsilon=0.
\end{align}

To bring the projector in the ``Hamiltonian'' form that allows us to read off the global and the dipole charges we multiply it by  $\frac{1+F_{12}\Gamma^{12}\sigma_3}{1+F_{12}^2}$, and define $\tan(\alpha) \equiv F_{12}$, obtaining
\begin{equation}
	\left(1+\cos(\alpha)^2\Gamma^{012}\sigma_1+\sin(\alpha)^2\Gamma^{01}+\cos(\alpha)\sin(\alpha)i\sigma_2\Gamma^{0}+\cos(\alpha)\sin(\alpha)\Gamma^{02}\sigma_3\right)\varepsilon=0.
\end{equation}
From this equation we can immediately identify the D2-brane and the momentum  global charges of the solution, as well as the dipole charges corresponding to F1 strings and D0 branes. It is straightforward to verify that the relations between these charges and the total energy are the same as for the BIon (\ref{sum-charges}, \ref{sum-charges-square}).

\bibliographystyle{utphys}   
\bibliography{Bibliographie}\end{document}